\documentclass[prd,eqsecnum,twocolumn,amsfonts,showpacs]{revtex4}

\usepackage{graphicx}

\usepackage{bm}

\def\fsl#1{\setbox0=\hbox{$#1$}           
   \dimen0=\wd0                                 
   \setbox1=\hbox{/} \dimen1=\wd1               
   \ifdim\dimen0>\dimen1                        
      \rlap{\hbox to \dimen0{\hfil/\hfil}}      
      #1                                        
   \else                                        
      \rlap{\hbox to \dimen1{\hfil$#1$\hfil}}   
      /                                         
   \fi}                                         %

\newcommand{\beq}{\begin{equation}}
\newcommand{\eeq}{\end{equation}}
\newcommand{\beqs}{\begin{eqnarray}}
\newcommand{\eeqs}{\end{eqnarray}}
\newcommand{\lsim}{\mathrel{\raisebox{-
.6ex}{$\stackrel{\textstyle<}{\sim}$}}}
\newcommand{\gsim}{\mathrel{\raisebox{-
.6ex}{$\stackrel{\textstyle>}{\sim}$}}}

\begin{document}

\title{Behavior of the $S$ Parameter in the Crossover Region Between Walking 
and QCD-Like Regimes of an SU($N$) Gauge Theory} 

\author{Masafumi Kurachi}

\author{Robert Shrock}

\affiliation{
C.N. Yang Institute for Theoretical Physics \\
State University of New York \\
Stony Brook, NY 11794}

\begin{abstract}

We consider a vectorial, confining SU($N$) gauge theory with a variable number,
$N_f$, of massless fermions transforming according to the fundamental
representation.  Using the Schwinger-Dyson and Bethe-Salpeter equations, we
calculate the $S$ parameter in terms of the current-current correlation
functions. We focus on values of $N_f$ such that the theory is in the crossover
region between the regimes of walking behavior and QCD-like (non-walking)
behavior.  Our calculations indicate that the contribution to $S$ from a given
fermion decreases as one moves from the QCD-like to the walking regimes.  The
implications of this result for technicolor theories are discussed.

\end{abstract}

\pacs{PACS No.}


\maketitle

\vspace{16mm}

\newpage
\pagestyle{plain}
\pagenumbering{arabic}

\section{Introduction}
\label{sec:Introduction}

The properties of a vectorial gauge theory as a function of the fermion content
are of fundamental importance.  Here we consider such a theory (in $(3+1)$
dimensions at zero temperature and chemical potential) with gauge group SU($N$)
and $N_f$ massless fermions transforming according to the fundamental
representation of this group.  For $N = N_c=3$ and $N_f=2$, this is an
approximation to actual quantum chromodynamics (QCD) with just the $u$ and $d$
quarks, since their current-quark masses are much smaller than the scale
$\Lambda_{QCD} \simeq 400$ MeV.  We restrict here to the range $N_f < (11/2)N$
for which the theory is asymptotically free.  An analysis using the two-loop
beta function and Schwinger-Dyson equation (reviewed below) leads to the
inference that for $N_f$ in this range, the theory includes two phases: (i) for
$0 \le N_f \le N_{f,cr}$ a phase with confinement and spontaneous chiral
symmetry breaking (S$\chi$SB), and (ii) for $N_{f,cr} \le N_f \le (11/2)N$ a
non-Abelian Coulomb phase with no confinement or spontaneous chiral symmetry
breaking. We shall refer to $N_{f,cr}$, the critical value of $N_f$, as the
boundary of the non-Abelian Coulomb (conformal) phase \cite{bz}.

For $N_f$ slightly less than $N_{f,cr}$, the theory exhibits an approximate
infrared (IR) fixed point. Let the SU($N$) running gauge coupling be denoted as
$g(\mu)$, where $\mu$ denotes the energy or momentum scale, and let
$\alpha(\mu) =\bar g(\mu)^2/(4\pi)$. As $\mu$ decreases from large values,
$\alpha(\mu)$ grows to be O(1) at a scale $\Lambda$, but increases only rather
slowly as $\mu$ decreases below $\Lambda$, so that there is an extended
interval in energy below $\Lambda$ where $\alpha$ is large, but slowly running
(``walking'').  In addition to its intrinsic field-theoretic interest, this
walking behavior is an essential ingredient of modern technicolor models of
dynamical electroweak symmetry breaking \cite{tc}, providing the requisite
enhancement of Standard-Model (SM) fermion masses \cite{wtc1}-\cite{chipt3}.
As $N_f$ approaches $N_{f,cr}$ from below, quantities with dimensions of mass
vanish continuously; i.e., the chiral phase transition separating phases (i)
and (ii) is continuous.

In this paper, we shall use solutions of the Schwinger-Dyson (SD) and
Bethe-Salpeter (BS) equations to compute a derivative of the difference of the
vector and axial-vector current-current correlation functions,
$\Pi'_{VV}(0)-\Pi'_{AA}(0)$.  Up to a multiplicative factor, this is the
coefficient $\bar L_{10}$ of one of the terms in the effective chiral
Lagrangian for the theory \cite{gl85,chipt}.  Moreover, in the context in which
one considers the SU($N$) theory as a technicolor (TC) model, with $N=N_{TC}$,
the above quantity is proportional to the correction to the $Z$ propagator due
to virtual electroweak-nonsinglet technicolor particles, often denoted as $S$
\cite{pt,ab}, \cite{pdg,lepewwg}. We focus on the crossover region between the
walking regime that occurs for $N_f \lsim N_{f,cr}$ and the QCD-like
(non-walking) regime that occurs for smaller $N_f$.

There are several motivations for this work. The quantity $S$, or equivalently
$\bar L_{10}$, is an intrinsic property of the SU($N$) theory, and it is of
interest to understand how this quantity depends on $N_f$. Further, our
calculations have important implications for technicolor models of dynamical
electroweak symmetry breaking \cite{wtc1}-\cite{chipt3}. For this application,
as noted, we identify the SU($N$) group with the technicolor gauge group.
Precision electroweak data \cite{pdg,lepewwg} determine allowed regions of
values of $S$ and the other $Z$ and $W$ propagator corrections denoted $T$ and
$U$, and yield a stringent constraint on the contributions from new particles
in technicolor models.  In order to assess the viability of these models, it is
necessary to have a reliable calculation of the contribution to $S$ from
technicolor particles.  A perturbative calculation of $S$ is not reliable since
the technifermions are strongly interacting on the relevant scale $\sim m_Z$.
Although it is possible to carry out a nonperturbative calculation for
technicolor theories that behave like scaled-up QCD (by using QCD data as input
for the relevant spectral functions), such theories are excluded since they
cannot produce sufficiently large standard-model fermion masses.  It is more
difficult to carry out a nonperturbative estimate of $S$ for technicolor models
that have walking behavior; such estimates have been presented in
Refs. \cite{ats}-\cite{iwts}.  These suggest that the contribution to $S$ per
technifermion electroweak doublet is reduced in the walking region. 
Meson masses and $f_P$ (the generalization of $f_\pi$) were
calculated in this walking regime in Ref. \cite{mmw}. (See also Ref. \cite{mme}
for the analogous calculations in QCD.)  A motivation for our
calculations in this paper is to gain further insight into the behavior of $S$
by studying its behavior in the crossover region between the walking and
QCD-like regimes. One of the reasons for concentrating on this crossover region
is that the theory still has an approximate infrared fixed point here, and
hence one can study the dependence of $S$ on $N_f$ without having to
introduce a model-dependent cutoff on the growth of the SU($N$) gauge coupling
in the infrared that was required in calculations of $S$ via the SD and BS
equations for small $N_f$ values, such as $N_f=3$ in QCD \cite{hys,pms}.  As
our previous calculations of meson masses and $f_P$ in this crossover region
showed, \cite{sg}, although it is a restricted interval in $N_f$ or
equivalently, the value of the infrared fixed point, it is sufficient large to
observe a significant change between walking and QCD-like behavior. 

This paper is organized as follows. Section II is devoted to a review some
background material concerning the beta function, approximate infrared fixed
point and walking behavior, and technicolor models. Section III contains
definitions of the current-current correlation functions and related spectral
functions together with the expression for $S$ in terms of these correlation
functions and an equivalent integral of spectral functions.  In this section we
also give the definition of $\bar L_{10}$ as a coefficient of a certain
operator in the low-energy effective chiral Lagrangian for the theory.  Section
IV explains how the current correlators are obtained from Bethe-Salpeter
amplitudes while Section V discusses the method of solution of the
Bethe-Salpeter equation.  Our results are presented in Section VI and their
implications for technicolor theories in Section VII.  An appendix discusses
some analytic approximations.

\section{Some Preliminaries} 
\label{sec:large_Nf}

In this section we review some background for our calculations.  For the theory
under consideration, with an SU($N$) gauge group and $N_f$ massless fermions in
the fundamental representation, the renormalization group (RG) equation for the
running coupling $\alpha(\mu)$ is 
\beq
\beta = \mu \frac{d \alpha(\mu)}{d \mu} = - \frac{\alpha(\mu)^2}{2\pi}
\left ( b_0 + \frac{b_1}{4\pi}\alpha + O(\alpha^2) \right ) \ ,
\label{eq:RGE_for_alpha}
\eeq
where $\mu$ is the momentum scale. The two terms listed are scheme-independent.
The next two higher-order terms have also been calculated but are
scheme-dependent; their inclusion does not significantly affect our results.
For the relevant case of an asymptotically free theory, $b_0 > 0$ so that an
infrared fixed point exists if and only if $b_1 < 0$.  This coefficient $b_1$
is positive for $0 \le N_f \le N_{f,IR}$, where $N_{f,IR}=(34N^3)/(13N^2-3)$,
and negative for larger $N_f$.  For $N=3$, $N_{f,IR} \simeq 8.1$
\cite{integer}.  The value of $\alpha$ at this IR fixed point, denoted
$\alpha_*$, is given by $\alpha_* = -4\pi b_0/b_1$.  Substituting the known
values of these terms \cite{b0,b1}, one has
\beq
\alpha_* = \frac{-4\pi(11N -2N_f)}{34N^2-13N N_f+3N^{-1}N_f} \ .
\label{alpha_irfp}
\eeq
Solving Eq. (\ref{alpha_irfp}) for $N_f$ in terms of $\alpha_*$ yields
\beq
N_f=\frac{2N^2[17N(\alpha_*/\pi)+22]}{(13N^2-3)(\alpha_*/\pi)+8N} \ .
\label{nfsol}
\eeq
As is evident from Eqs. (\ref{alpha_irfp}) and (\ref{nfsol}), $\alpha_*$ is a
monotonically decreasing function of $N_f$ and equivalently $N_f$ is a
monotonically decreasing function of $\alpha_*$, for $N_{f,IR} \le N_f \le
(11/2)N$. 

To study the dependence of $S$ on $N_f$, what we actually vary is the value of
the approximate IR fixed point $\alpha_*$, which depends parametrically on
$N_f$.  For definiteness, we shall take $N=3$; however, as will be seen,
$N$ only enters indirectly, via the dependence of the value of the infrared
fixed point $\alpha_*$ (Eq. (\ref{alfcrit}) below) on $N_c$.  Hence, our
findings may also be applied in a straightforward way, with appropriate changes
in the value of $\alpha_*$, to an SU($N$) gauge theory with a different value
of $N$.

In the one-gluon exchange approximation, the Schwinger-Dyson gap equation for
the inverse propagator of a fermion transforming according to the
representation $R$ of SU($N$) has a nonzero solution for the dynamically
generated fermion mass, which is an order parameter for spontaneous chiral
symmetry breaking, if $\alpha \ge \alpha_{cr}$, where $\alpha_{cr}$ is given by
\beq
\frac{3 \alpha_{cr} C_2(R)}{\pi} = 1,
\label{alfcritcondition}
\eeq
and $C_2(R)$ denotes the quadratic Casimir invariant for the representation $R$
\cite{casimir}.  Using
\beq
C_2(fund.) \equiv C_{2f}=\frac{N^2-1}{2N}
\label{c2f}
\eeq
for the fundamental representation yields
\beq
\alpha_{cr} = \frac{2\pi N}{3(N^2-1)} \ .
\label{alfcrit}
\eeq
For the case $N=3$ that we use for definiteness here, Eq. (\ref{alfcrit})
gives $\alpha_{cr} = \pi/4 \simeq 0.79$.  To estimate $N_{f,cr}$, one solves
the equation $\alpha_* = \alpha_{cr}$, yielding the result \cite{chipt3}
\beq
N_{f,cr} = \frac{2N(50N^2-33)}{5(5N^2-3)} \ .
\label{nfcr}
\eeq
For the values $N=3$ and $N=2$ this gives $N_{f,cr} \simeq 11.9$ and $N_{f,cr}
\simeq 7.9$, respectively.  These estimates are only rough, in view of the
strongly coupled nature of the physics. Effects of higher-order gluon 
exchanges have been studied in Ref. \cite{alm}.  In principle, lattice gauge
simulations should provide a way to determine $N_{f,cr}$, but the groups that
have studied this have not reached a consensus \cite{lgt}.

If $\alpha_*$ is less than the critical value, $\alpha_{cr}$, for a bilinear
fermion condensate to form, the above IR fixed point is exact, with the
coupling $\alpha$ approaching $\alpha_*$ from below as the energy scale $E$
decreases from large values to zero.  Let us denote the fermions as $f_i^a$
with $a=1,...,N$ and $i=1,...,N_f$.  If $\alpha_* > \alpha_{cr}$, as $E$
descends from large values, the coupling $\alpha$ eventually exceeds the above
critical value, the fermion condensates $\langle \sum_{a=1}^{N} \bar f_{a,i}
f^a_i \rangle$ (with no sum on $i$) form, and are equal for each flavor
$i=1,...,N_f$ (with electroweak interactions negligibly small relative to the
SU($N$) interaction).  Accordingly, the global ${\rm SU}(N_f)_L \times {\rm
SU}(N_f)_R \times {\rm U}(1)_V$ symmetry (where U(1)$_V$ is fermion number) is
broken to generalized isospin times fermion number, ${\rm SU}(N_f)_V \times
U(1)_V$.

Associated with this, the fermions pick up dynamical masses $\Sigma$ and are
integrated out as the energy scale decreases below $\Sigma$.  Hence, in this
case, the IR fixed point is only approximate since in the effective field
theory for energies below $\Sigma$, the form of the beta function is that for
the pure gauge theory, $N_f=0$.  The case where $\alpha_*$ is close to, and
slightly larger than, $\alpha_{cr}$, yields walking behavior.  In the strong
walking regime, the dynamical fermion mass, and also hadron masses are
exponentially smaller than the scale $\Lambda$ at which the coupling first
becomes O(1) as the energy scale decreases from large values.  Although our SD
and BS equations are semi-perturbative, the analysis is self-consistent in the
sense that our $\alpha_{cr}$ really is the value at which, in our
approximation, one passes from the confinement phase to the non-Abelian Coulomb
phase, and our values of $\alpha$ do span the interval over which there is a
crossover from walking to QCD-like (i.e., non-walking) behavior.

As is evident from the above results, decreasing $N_f$ below $N_{f,cr}$ has the
effect of increasing $\alpha_*$ and thus moving the theory deeper in the phase
with confinement and spontaneous chiral symmetry breaking, away from the
boundary with the non-Abelian Coulomb phase.  This is the key parametric
dependence that we shall use for our study.  In Refs. \cite{mmw,pmsw} the range
of $\alpha_*$ used for the calculation of meson masses was chosen to be $0.89
\le \alpha_* \le 1.0$, an interval where there is pronounced walking behavior.
For the case $N=3$ considered in Ref. \cite{pmsw} and here, given the
above-mentioned value, $\alpha_{cr}=\pi/4$, it follows that this lower limit,
$\alpha_*=0.89$, is about 12 \% greater than this critical coupling.  The
reason for this choice of lower limit on $\alpha_*$ was that the calculation of
$S$ involves very strong cancellations as $\alpha_* - \alpha_{cr} \to 0^+$,
rendering it progressively more and more difficult to obtain accurate numerical
results in this extreme walking limit.  For our present study of $S$ we
consider an interval extending to larger couplings, from $\alpha_*=1.0$ to
$\alpha_*=1.8$.  Our upper limit is chosen in order for the ladder
approximation used in our solutions of the Schwinger-Dyson and Bethe-Salpeter
equations to have reasonable reliability.  From Eq. (\ref{nfsol}) it follows
that $\alpha_*=0.89$ corresponds to $N_f=11.65$, about 2 \% less than
$N_{f,cr}$.  For a coupling as large as $\alpha_* = 1.8$, the semi-perturbative
methods used to derive eqs. (\ref{alpha_irfp}) and (\ref{nfsol}) are subject to
large corrections from higher-order perturbative, and from nonperturbative,
contributions; recognizing this, the above upper limit of $\alpha_*$
corresponds to $N_f \simeq 10.3$, a roughly 13 \% reduction from
$N_{f,cr}=11.9$.  Although this shift appears to be by only a modest amount
when expressed in terms of $N_f$, in terms of $\alpha_*$ it is a factor of
two, and our calculations in Ref. \cite{sg} showed a dramatic change in the
values of meson mass ratios and $f_P/\Lambda$ in this range, with these values
changing from their walking limits toward QCD-like values.  Hence we anticipate
that this range can be sufficient to study the shift in the value of $S$, and
our results confirm this.

If $N_f > N_{f,cr}$, i.e., $\alpha_* < \alpha_{cr}$ so that this IR fixed point
of the two-loop RG equation is exact, then, denoting $b \equiv b_0/(2\pi)$, the
solution to this equation can be explicitly
written~\cite{Gardi,ExplicitSolution} in the entire energy region as
\beq
\alpha(\mu)=\alpha_\ast \left[\ W(e^{-1}(\mu/\Lambda)^{b \alpha_*})+1 \ 
\right]^{-1},
\label{Lambert}
\eeq
where $W(x) = F^{-1}(x)$, with $F(x) = x e^x$, is the Lambert $W$
function, and $\Lambda$ is a RG-invariant
scale defined by~\cite{chipt2}
\beq
 \Lambda \ \equiv\  \mu \ \exp \left[ -\frac{1}{b} \left \{ 
\frac{1}{\alpha_\ast}
\ln\left( \frac{\alpha_\ast - \alpha(\mu)}{\alpha(\mu)} \right)
        + \frac{1}{\alpha(\mu)} \right \} \right] .
\label{eq:Lambda}
\eeq 
Now since we are studying the confined phase with $N_f < N_{f,cr}$, ($\alpha_*
> \alpha_{cr}$) with spontaneous chiral symmetry breaking, $\alpha_*$ is only
an approximate, rather than exact, IR fixed point.  Hence, the solution
(\ref{Lambert}) is only applicable in an approximate manner to our case; for
momenta much less than the dynamical fermion mass $\Sigma$, the fermions
decouple, and in this very low-momentum region, with the fermions integrated
out, the resultant $\alpha$ would increase above the value $\alpha_*$ at the
approximate IR fixed point.  However, since $\Sigma \ll \Lambda$ in a walking
or near-walking theory, it follows that this lowest range of momenta makes a
small contribution to the relevant integrals to be evaluated in our
calculations.  Hence, over most of the integration range for these integrals
where the coupling $\alpha$ is large, it is approximately constant and equal to
its fixed-point value, $\alpha_*$ (see Fig. 2 of Ref. \cite{mmw}).  This means
that one can use, as a reasonable approximation, the expression 
\beq
  \alpha(\mu) = \alpha_\ast \, \theta(\Lambda-\mu),
\label{run_approx}
\eeq
where $\theta$ is the step function.  (This is the same approximation used in
Refs. \cite{sg,mmw}.)  Thus, in the walking region and the adjacent
crossover region, the calculations have the advantage that one can avoid having
to introduce an artificial cutoff on the growth of $\alpha$ in the infrared, in
contrast to the situation for smaller $N_f$, where the walking behavior
disappears and this cutoff is necessary. 

Since an important application of our results is to technicolor models, we
briefly mention some relevant features of these models.  As noted, the
technicolor gauge theory has a gauge group SU($N_{TC}$) and an asysmptotically
free coupling that gets large at the TeV scale \cite{tc}.  It contains a set of
massless, vectorially coupled technifermions.  The left-handed components of
these fermions transform as doublets under SU(2)$_L$.  The spontaneous chiral
symmetry breaking and formation of a bilinear technifermion condensate breaks
the electroweak symmetry from ${\rm SU}(2)_L \times {\rm U}(1)_Y$ to
U(1)$_{em}$, producing masses for the $W$ and $Z$ given to leading order by
$m_W^2 = m_Z^2 \cos^2 \theta_W = (g^2/4)N_{D,TF} F_{TC}^2$, where $F_{TC}$ is
the technicolor analogue of $f_\pi$. In order to give masses to the
Standard-Model fermions (which are technisinglets), it is necessary to embed
technicolor in a larger, extended technicolor (ETC) theory \cite{etc,tcm}, with
interactions that transform technifermions to the Standard-Model fermions and
vice versa.  To satisfy constraints from flavor-changing neutral-current (FCNC)
processes, the ETC vector bosons that mediate generation-changing transitions
must have large masses, ranging from a few TeV to $10^3$ TeV.  For our present
study, concerned with $S$, we concentrate on the technicolor theory at the
scale of a few hundred GeV, with the ETC gauge bosons integrated out.

We focus here on models in which the technifermions transform according to the
fundamental representation of the SU($N_{TC}$) gauge group.  Two simple
examples are the so-called one-doublet and one-family technicolor models.  A
one-doublet TC model has $N_f=2$ technifermions, denoted $U$ and $D$, whose
chiral components transform according to
\beqs
& & F_L = {U_L \choose D_L} \ : \ (N_{TC},1,2)_{0,L}, \cr\cr
& & U_R \ : \ \ (N_{TC},1,1)_{1,R}, \cr\cr
& & D_R \ : \ \ (N_{TC},1,1)_{-1,R} \ . 
\label{1d}
\eeqs
where the numbers in parentheses refer to the dimensions of the representations
of ${\rm SU}(N_{TC}) \times {\rm SU}(3)_c \times {\rm SU}(2)_L$ and the
subscripts refer to the weak U(1)$_Y$ hypercharges.  The value $N_{TC}=2$ has
been preferred in recent TC/ETC model-building \cite{at94,as} for several
reasons, including the fact that it (i) minimizes technicolor contributions to
the $S$ parameter, (ii) can naturally produce a walking theory in a one-family
model (see below), and (iii) makes possible a mechanism to explain
light neutrino masses \cite{as}.  The one-doublet TC model has one
SU(2)$_L$ doublet of technifermions for each technicolor index, which we
express as $N_{D,TF}=1$, and hence a total number of SU(2)$_L$ doublets
of technifermions equal to $N_{D,tot}=N_{D,TF}N_{TC}=N_{TC}$.  The TC sector
with just these $N_f=2$ technifermions would not exhibit walking behavior, but
one can add SM-singlet technifermions to produce a theory that does have such
behavior \cite{ts}.  In a one-family TC model the technifermions transform as
\beqs
& & Q_L: \ (N,3,2)_{1/3,L} \cr\cr
& & u_R: \ (N,3,1)_{4/3,R} \cr\cr
& & d_R: \ (N,3,1)_{-2/3,R} \cr\cr
& & L_L: \ (N,1,2)_{-1,L} \cr\cr
& & N_R: \ (N,1,1)_{0,R} \cr\cr
& & E_R: \ (N,1,1)_{-2,R}
\label{1fam_quarks}
\eeqs
Hence, this type of technicolor models contains $N_f=2(N_c+1)=8$
technifermions. As is evident from Eq.  (\ref{nfcr}), with $N=N_{TC}=2$, the
value $N_f=8$ is close to the value $N_{f,cr}$ and hence, to within the
accuracy of the two-loop beta function analysis, this technicolor model can
naturally exhibit walking behavior.  Reverting to general $N=N_{TC}$ for our
discussion, this one-family technicolor model thus has $N_{D,TF}=(N_c+1)=4$
\ SU(2)$_L$ doublets for each technicolor index, and hence a total of 
$N_{D,tot.}=4N_{TC}$ SU(2)$_L$ doublets of technifermions. 

Because of the spontaneous chiral symmetry breaking in the technicolor theory,
the technifermions pick up dynamical masses $\Sigma_{TC}$ proportional to
$F_{TC} N_{TC}^{-1/2}$, where we have included the $N_{TC}$-dependent factor
that would be present in the large-$N_{TC}$ limit, since $f_P$ and
$\Sigma_{TC}$ scale, respectively, like $N_{TC}^{1/2}$ and $N_{TC}^0$ in this
limit.  For the one-doublet and one-family TC models, $F_{TC} \simeq 250$ and
125 GeV, respectively.  In QCD, the constituent quark mass $\Sigma \simeq 3.5
f_\pi$, and one expects a roughly similar ratio in TC theories (see Fig. 3 in
our previous work \cite{sg}).  Since the SM gauge couplings are small at the
technicolor scale, different technifermions are expected to have roughly
degenerate dynamical masses, and the contributions of the techniquark and
technilepton doublets to one-loop corrections to the $Z$ propagator are
approximately equal.


\section{Expression for $S$ in Terms of Current-Current Correlation 
Functions}
\label{sec:Spectral}

As a measure of corrections to the $Z$ propagator arising from heavy particles
in theories beyond the Standard Model, $S$ was originally defined as 
\cite{pt}
\beq
S = \frac{4 s_W^2 c_W^2}{\alpha_{em}(m_Z)} \left. \frac{d \Pi^{(NP)}_{ZZ}(q^2)}
{dq^2} \right\vert_{q^2 = 0} 
\label{s}
\eeq
where $s_W^2 = 1-c_W^2 = \sin^2\theta_W$, evaluated at $m_Z$ and the
superscript $NP$ refers to the fact that the definition includes new physics
beyond the Standard Model.  In the case of technicolor, the technifermions 
are taken to have zero masses; because of the spontaneous chiral symmetry 
breaking in the TC theory, they pick up dynamical masses $\Sigma_{TC}$ of order
the technicolor scale.  More recent analyses of precision electroweak data
define $S$ slightly differently, replacing the derivative at $q^2=0$ by a
discrete difference (in the $\overline{MS}$ scheme) \cite{pdg} 
\beq
S_{PDG} = \frac{4 s_W^2 c_W^2}{\alpha_{em}(m_Z)} \Bigg [ 
\frac{\Pi^{(NP)}_{ZZ}(m_Z^2) - \Pi^{(NP)}_{ZZ}(0)}{m_Z^2} \Bigg ] \ . 
\label{spdg}
\eeq
The difference between these definitions is small if the heavy physics scale
$\Sigma_{TC}$ satisfies $(2\Sigma_{TC}/m_Z)^2 \gg 1$, as is the case in the
TC models considered here.

For our purposes it will be convenient to use the original definition, Eq.
(\ref{s}).  The implications of our results for technicolor theories would be
essentially the same if we used the expression (\ref{spdg}).  With either
definition, since a one-loop heavy fermion correction to the $Z$ propagator has
a prefactor $(g^2+g'^2)/(16\pi^2)$, where $g$ and $g'$ are the respective
SU(2)$_L$ and U(1)$_Y$ gauge couplings, and since $(g^2+g'^2)/(16\pi^2)=
\alpha_{em}/(4\pi s_W^2 c_W^2)$, the prefactor $4 s_W^2 c_W^2/\alpha_{em}(m_Z)$
in the definition of $S$ cancels out the leading dependence on the electroweak
gauge couplings (evaluated at the scale $m_Z$), yielding a quantity that
depends on the intrinsic properties of the strongly coupled SU($N$) gauge
theory.

Now, suppressing the SU($N_{TC}$) gauge index, we write the fermions as a
vector, $\psi = (\psi_i,...,\psi_{N_f})$. We then define vector and the
axial-vector currents as
\beqs
   V_\mu^a(x) &=& \bar\psi(x) T^a \gamma_\mu \psi(x) \cr\cr
   A_\mu^a(x) &=& \bar\psi(x) T^a \gamma_\mu \gamma_5\psi(x) \, ,
  \label{def:currents}
\label{vacurrents}
\eeqs
where the $N_f \times N_f$ matrices $T^a$ ($a=1,..., N_f^2-1$) are the
generators of $SU(N_f)$ with the standard normalization ${\rm Tr}(T^a T^b) =
\frac{1}{2}\delta^{ab}$.  In terms of these currents, the two-point
current-current correlation functions $\Pi_{VV}$ and $\Pi_{AA}$ are defined via
the equations
\beqs 
&& i\int d^4x\ e^{i q \cdot x}\ \langle 0|T(J_\mu^a(x) J_\nu^b(0))|0 \rangle \
\cr\cr
&=& \delta^{ab}\left(\frac{q_\mu q_\nu}{q^2}-g_{\mu\nu}\right)\Pi_{JJ}(q^2) \ ,
\label{pijjdef}
\eeqs
where $J_\mu^a(x) = V_\mu^a(x), A_\mu^a(x)$. With the above normalization of
$T^a$, $\Pi(q^2)$ measures the contributions to the time-ordered product in
Eq. (\ref{pijjdef}) per fermion.  Given that, to a good approximation,
different technifermion doublets contribute equally to $S$, it is natural to
define a reduced quantity, $\hat S$, that represents the contribution to $S$
from each such pair, viz.,
\beq
\hat S = \frac{S}{N_D} \ . 
\label{shat}
\eeq

Then, in terms of the current-current correlation functions defined above,
$S$, as defined in Eq. (\ref{s}), is given by 
\beq
\hat S = 4 \pi \left. \frac{d}{d q^2} \left[ \Pi_{VV}(q^2) - \Pi_{AA}(q^2)
   \right] \right\vert_{q^2 = 0}\, ,
 \label{scor}
\eeq
It is convenient to define the compact notation 
\beq
\Pi_{V-A}(q^2) \equiv \Pi_{VV}(q^2) - \Pi_{AA}(q^2) \ . 
\label{piva}
\eeq

As is evident from Eq. (\ref{scor}), one may also consider $S$ in a different
context, namely that of an abstract vectorial SU($N$) gauge theory with $N_f$
massless fermions transforming according to the fundamental representation of
this group, and with all other interactions much weaker in strength than the
SU($N$) gauge interaction.  In this case, in contrast to technicolor models,
where $N_f$ is even (since the technifermions have left-handed components
forming SU(2)$_L$ doublets), $N_f$ can be even or odd (being restricted to be
less than $N_{f,cr}$ so that the theory is in the confinement phase with
spontaneous chiral symmetry breaking).  Here, one could naturally define the
contribution to $S$ from each fermion individually, namely, $S/N_f$.  However,
since our main application will be to technicolor, we shall continue, as in
Ref. \cite{pmsw}, to present our results in terms of $\hat S$.

Because of the asymptotic freedom of the SU($N$) theory, for Euclidean $q^2$
much larger than $\Lambda^2$, dimensional considerations imply that,
asymptotically, $\Pi_{V-A}(q^2) \simeq \langle \bar\psi\psi \rangle^2/q^4$ up
to logs arising from anomalous dimensions.  Combining this property with the 
analytic properties of $\Pi_{V-A}(q^2)$, one can write the following 
dispersion relation, with $t \equiv q^2$:
\beq
\Pi_{V-A}(t) = \frac{1}{\pi} \int_0^\infty ds \, 
\frac{ Im(\Pi_{V-A}(s))}{s-t-i\epsilon}
\label{disprel}
\eeq
The dimensionless spectral function $\rho_J(s)$ is defined as 
\beq
\rho_J(s) \equiv \frac{Im(\Pi_{JJ}(s))}{\pi s}
\label{rho}
\eeq
This spectral function encodes information about the hadronic states produced
by the current $J$.  In terms of these spectral functions, one then has
\beq
\hat S = 4 \pi \int_0^\infty  \frac{ds}{s} \bigg [ \rho_V(s)-\rho_A(s) \bigg ] 
\label{w0}
\eeq
An early discussion of the integral on the right-hand side of Eq. (\ref{w0})
and its connection to $\pi^+ \to \ell^+ \nu_\ell \gamma$ decay appeared in
\cite{dmo}.  It is of interest to comment on the two equivalent expressions,
Eqs. (\ref{scor}) and (\ref{w0}) for $\hat S$.  The first of these,
Eq. (\ref{scor}), expresses $\hat S$ in terms of the slope of $\Pi_{V-A}(q^2)$
at $q^2=0$.  In contrast, the second, Eq. (\ref{w0}) expresses it as an
integral of $1/s$ times the difference of the physical spectral functions for
the vector and axial vector currents in the timelike region, which depends on
the spectrum of hadronic states and the strengths of their couplings to these
currents.  Thus, naively, one might think that these two expressions depend on
rather different properties of the SU($N$) theory.  The fact that they are
actually equivalent is a consequence of the analytic properties of
$\Pi_{JJ}(q^2)$ which are used, via the Cauchy theorem, to derive the resultant
dispersion relation.  Thus, although it is evaluated at a single point, the
derivative $\Pi_{V-A}'(0)$ ``knows'' about the full hadronic spectrum,
including the presence or absence of walking.

In passing, we note that the spectral functions directly in terms of the 
products of currents, as 
\beqs
& & i\int d^4x \ e^{i q \cdot x} \ \langle 0|T(V_\mu^a(x)V_\nu^b(0))|0\rangle  
\cr\cr
&=& -\delta^{ab} \int_0^\infty \, ds \, 
\frac{\rho_V(s)(sg_{\mu\nu}-q_\mu q_\nu)}{s-q^2-i\epsilon} 
\label{rhov}
\eeqs
and
\beqs
&& i \int d^4x \ e^{i q \cdot x} \ \langle 0|T(A_\mu^a(x)A_\nu^b(0))|0\rangle  
\cr\cr 
&=& -\delta^{ab}\bigg [ \int_0^\infty \, ds \,
\frac{\rho_A(s)(sg_{\mu\nu}-q_\mu q_\nu)}{s-q^2-i\epsilon} + 
\frac{q_\mu q_\nu}{q^2} f_P^2 \ \bigg ] \ , \cr\cr
& & 
\label{rhoa}
\eeqs
where one explicitly separates out the contribution from the ($J=0$ part 
of) $\rho_A(s)$ at $s=0$ due to the Nambu-Goldstone bosons (NGB's). In
Eq. (\ref{rhoa}), the quantity $f_P$ (where $P$ stands for ``pseudoscalar'') is
the $N_f$-flavor generalization of $f_\pi$ defined by the transition matrix 
element
\beq
   \langle 0 \vert A_\mu^a (0) \vert \pi^b(q) \rangle 
   = i q_\mu f_P \delta^{ab}\, ,
\label{fp}
\eeq 
with $a,b=1,2,\ldots,N_f^2-1$.  In actual QCD, the chiral symmetry is
explicitly broken by the $u$ and $d$ current-quark masses (and also by
electroweak interactions), so that the pions decay, and, in particular, the
dominant weak decay of the $\pi^+$, $\pi^+ \to \mu^+ \nu_\mu$ has a rate
proportional to $f_\pi^2$. Thus, $f_P$ might be called the generalized
Nambu-Goldstone decay constant, but we will avoid this term, since in our basic
SU($N$) theory with other interactions turned off, these Nambu-Goldstone bosons
are exactly massless and do not decay.  In the chiral limit of QCD, with
$m_u=m_d=0$, it has been estimated that $(f_\pi)_{ch.lim.}/f_\pi \simeq 0.935$
\cite{gl85}, so that, with the physical value $f_\pi=92.4 \pm 0.3$, one infers
that $(f_\pi)_{ch.lim.} \simeq 86$ MeV (with a theoretical uncertainty of
several per cent from the chiral extrapolation).  This slight decrease will not
be important for our work.

The constant $f_P$ may be calculated by first solving the Schwinger-Dyson
equation for the momentum-dependent dynamical fermion mass $\Sigma(p)$ and then
substituting this into the Pagels-Stokar relation \cite{psrel},
\beq
 f_P^2 \, = \, \frac{N_c}{4\pi^2} \int_0^\infty y \, dy \,
       \frac{\Sigma^2(y) \, - \,  \frac{y}{4} \,
\frac{d}{dy} \left[ \Sigma^2(y) \right] }
       { [ y \ +\  \Sigma^2(y) \ ]^2}  \ .
\label{psrel}
\eeq
Calculations using this method~\cite{hys,mmw,sg} have shown (for $N=N_c=3$)
that as $N_f$ increases from the value $N_f=2$ toward $N_{f,cr}$, the
generalized quantity $f_P$ decreases, as is expected, since $f_P$ is an order
parameter for spontaneous chiral symmetry breaking.  Furthermore, in the strong
walking limit $N_f \nearrow N_{f,cr}$, i.e., $\alpha_* \searrow \alpha_{cr}$,
it has been found that $f_P$ vanishes like \cite{my,miranskybook}
\beq
f_P = c_f \Lambda \, \exp
\bigg [ -\pi \Big ( \frac{\alpha_*}{\alpha_{cr}} - 1 \Big )^{-1/2} \bigg ] \ ,
\label{sigsol}
\eeq
where $c_f$ is a constant.  In Ref. \cite{sg} we calculated $f_P$ in the
crossover region between this extreme walking limit and smaller values of
$N_f$, corresponding to larger values of $\alpha_*$, closer to QCD with $N_f=2$
or 3 and found a dramatic growth in $f_P/\Lambda$, approaching values nearer to
QCD, as expected.

Because of the asymptotic freedom and consequent asymptotic decay of
$\Pi_{V-A}(q^2)$ for large Euclidean $q^2$ and the fact that the fermions are
massless in the underlying SU($N$) theory, the following integral relations
(first and second Weinberg sum rules) hold \cite{wsumrule}-\cite{dg}:
\beq
\int_0^\infty \, ds \, \bigg [\rho_V(s)-\rho_A(s) \bigg ] = 
\Pi_{V-A}(0) = f_P^2 
\label{w1}
\eeq
and
\beq
\int_0^\infty \, ds \, s \,  \bigg [\rho_V(s) - \rho_A(s) \bigg ] = 0 \ . 
\label{w2}
\eeq
Clearly, Eqs. (\ref{w0}), (\ref{w1}), and (\ref{w2}) are of similar form,
viz., integrals of $\rho_V(s)-\rho_A(s)$ weighted, respectively by $s^{-1}$, 
1, and $s$.  Because of these different factors, the contributions to
the integral (\ref{w0}) for $\hat S$ are weighted toward smaller 
$s$ values, while the contributions to the integral (\ref{w2}) are weighted
toward larger $s$ values, as compared with those for the integral in Eq. 
(\ref{w1}). 

Finally, in the context of an abstract SU($N$) gauge theory (with other
interactions turned off), it is worthwhile to mention the connection between
$S$ and the coefficient denoted $\bar L_{10}$.  To explain this connection, we
recall the definition of $\bar L_{10}$. Provided that $N_f \ge 2$, the theory
contains $N_f^2-1$ massless Nambu-Goldstone bosons.  (For this discussion, we
turn off electroweak interactions completely; for our application to
technicolor, we do not turn them off, and hence three of the would-be
Nambu-Goldstone bosons become the longitudinal modes of the $W^\pm$ and
$Z$. Furthermore, in the TC/ETC context, other would-be NGB's gain masses from
color and ETC interactions that explicitly break the ${\rm SU}(N_f)_L \times 
{\rm SU}(N_f)_R$ global chiral symmetry.) 
A useful description of the low-energy physics for energies $E \ll 4 \pi f_P$
is then provided by a chiral Lagrangian (see. e.g., \cite{gl85} and references
to earlier work therein).  This is a function of the chiral fields
\beq
U = \exp \left ( \frac{2 i \sum_{a=1}^{N_f^2-1} \pi^a T^a}{f_P} \right ) \ . 
\label{uchiral}
\eeq
For example, for QCD with just the two light quarks $u$ and $d$, $U$ would have
the form $U=e^{i \pi \cdot \tau/f_\pi}$.  Now let us denote the elements of the
global symmetry groups ${\rm SU}(N_f)_L$ and ${\rm SU}(N_f)_R$ as $U_L$ and
$U_R$.  Then under a transformation by an element of the chiral symmetry group
${\rm SU}(N_f)_L \times {\rm SU}(N_f)_R$, $U \to U_L U U_R^\dagger$.  One can
also formally introduce external gauge fields (contracted with the respective
generators of ${\rm SU}(N_f)_L$ and ${\rm SU}(N_f)_R$), $L_\mu$ and $R_\mu$.
These have the transformation properties $L_\mu \to U_L L_\mu U_L^\dagger - i
(\partial_\mu U_L) U_L^\dagger$ and $R_\mu \to U_R R_\mu U_R^\dagger - i
(\partial_\mu U_R) U_R^\dagger$.  In terms of these, one constructs the
covariant derivative $D_\mu U = \partial_\mu U - i L_\mu U + i U R_\mu$.  One
also defines external field strengths $(F_L)_{\mu\nu}$ and $(F_R)_{\mu\nu}$ in
the usual manner.  The lowest-order term in this effective Lagrangian is then
\beq
 \frac{f_P^2}{4} \, {\rm Tr} \left [ (D_\mu U)^\dagger (D^\mu U) \right ] \ . 
\label{lowestterm}
\eeq
In terms of these quantities, $L^{(r)}_{10}$ is the coefficient of the term 
\beq
O_{10} = {\rm Tr}(U^\dagger (F_L)_{\mu\nu} U (F_R)^{\mu\nu}) , 
\label{o10}
\eeq
where the superscript $(r)$ refers to the renormalized quantity. From this, one
obtains a quantity which is constructed to be independent of the
renormalization scale $\mu$ \cite{gl85,chipt} 
\beq
\bar L_{10} = L_{10}^{(r)}(\mu) + \frac{1}{192\pi^2} 
\Bigg [ \ln \left ( \frac{m_\pi^2}{\mu^2} \right ) + 1 \Bigg ]
\label{l10bar}
\eeq
The relation with $S$ is then 
\beq
S = -16 \pi \bar L_{10} \ . 
\label{slrel}
\eeq
Note that although $\bar L_{10}$ requires $N_f \ge 2$ to be well-defined, it
does not require $N_f$ to be even.  For the specific case of QCD, fits to
experimental data, including, in particular, the radiative decay $\pi^+ \to e^+
\nu_e \gamma$, yield the value \cite{gl85,hyrev}
\beq
S = 0.33 \pm 0.04 \ . 
\label{svalue}
\eeq

To the extent that this is dominated by the two light quarks $u$ and $d$, one
has $N_f=2$, and hence $N_D=1$, so that the measured value of $S$ for QCD is
also equal to $\hat S$.  The fact that the light-quark vector and axial-vector
mesons $\rho$ and $a_1$ largely saturate the expression for $S$,
Eq. (\ref{w0}), is consistent with this conclusion.  An approximate calculation
of $\hat S$ has been performed using the ladder approximation to the
Schwinger-Dyson and (inhomogeneous) Bethe-Salpeter equations for QCD ($N=3$)
with $N_f=2$ quarks of negligible mass \cite{hys}.  Studies have also been done
for the case where one neglects the strange quark mass $m_s$, i.e., $N=3$,
$N_f=3$ \cite{hys,pms}.  Since for either of these values of $N_f$ the beta
function of the QCD theory does not exhibit an infrared fixed point, it is
necessary to cut off the growth of the strong coupling.  For typical cutoffs,
it was found that the SD-IBS calculations tended to yield slightly too large a
value of $\hat S = S$, namely $\hat S \simeq 0.45 - 0.5$ \cite{hys,pms}, rather
than a value in the $1\sigma$ experimental range $0.29 \lsim S \lsim 0.37$.
This suggests that in the SD-IBS approach, used for a vectorial confining
SU($N$) gauge theory with small values of $N_f$ such that the theory has no
perturbative IR fixed point, with a typical IR cutoff on the coupling, tends to
overestimate $S$ by about a factor of 1.4.  Since the calculation of $S$ in QCD
is a problem in strongly coupled, nonperturbative physics, and the
calculational method that was used is only approximate (neglecting, for
example, instanton contributions), one should probably not be surprised that it
does not precisely reproduce the measured value of $S$.

\section{Current-Current Correlation Functions in Terms of Bethe-Salpeter 
Amplitudes}
\label{sec:correlator-BS}
 
In this section, we explain how the current-current correlation functions are
obtained from the Bethe-Salpeter amplitudes, which will be calculated via the
inhomogeneous Bethe-Salpeter (IBS) equation \cite{bs}-\cite{marisroberts03},
\cite{miranskybook}.  These Bethe-Salpeter amplitudes $\chi^{(J)}$, where $J=V$
or $A$, are essentially form factors, whose behavior in the timelike region
describes the coupling of the given current to physical hadronic bound states
that can be produced by this current, with an analytic continuation into the
spacelike region.  Here we will only need these amplitudes for the spacelike
region $q^2 < 0$ and at the point $q^2=0$. The amplitudes may be defined in
terms of the three-point vertex function as
\beqs 
& & \delta_j^k \left( T^a \right)_{f}^{f'} \int \frac{d^4 p}{(2 \pi)^4} 
\, e^{- i p \cdot r} \, \chi^{(J)}_{\alpha \beta}(p;q,\epsilon) = \cr\cr
& & = \epsilon^\mu \int d^4 x \, e^{i q \cdot x} 
\langle 0 \vert T(\psi^k_{\alpha f}(r/2) \ \bar\psi_{jf'\beta}(-r/2)\
J_\mu^a(x)) | 0 \rangle, \cr\cr
& & 
\label{eq:three-point}
\eeqs
where $(f,f')$, $(j,k)$ and $(\alpha, \beta)$ are, respectively, the flavor,
gauge, and spinor indices.  Closing the fermion legs of the above
three-point vertex function and taking the limit $r \rightarrow 0$, one can
express the current correlator in terms of the Bethe-Salpeter amplitude as
\beqs
& &  \Pi_{JJ}(q^2) = \cr\cr
& & \frac{1}{3} \Bigg ( \frac{N}{2} \Bigg ) \sum_{\epsilon} 
 \int \frac{d^4 p}{i (2\pi)^4} 
{\rm Tr} \left[ \left(\epsilon \cdot G^{(J)}\right) 
   \chi^{(J)}(p;q,\epsilon) \right] \cr\cr
& & 
 \label{eq:Pi_JJ}
\eeqs
where \cite{gamcon}
\beq
G_\mu^{(V)} = \gamma_\mu,\ \ \ \ G_\mu^{(A)} = \gamma_\mu \gamma_5,
\eeq
and an average has been taken over the polarizations, so that $\Pi_{JJ}(q^2)$
does not depend on the polarization $\epsilon$.

We expand the Bethe-Salpeter amplitude $\chi_{\alpha
\beta}^{(J)}(p;q,\epsilon)$ in terms of a complete bispinor basis with basis
elements $\Gamma^{(J)}_i$ and the invariant amplitudes $\chi^{(J)}_i$ as
\beq
   \left[ \chi^{(J)}(p;q,\epsilon) \right]_{\alpha \beta} 
   =\ \sum_{i=1}^{8} \left[ \Gamma_i^{(J)}(p;\hat{q},\epsilon)
   \right]_{\alpha \beta} \chi^{(J)}_{i} (p;q) ,
 \label{eq:chi-expanded}
\eeq
where we define
\beq
Q^2 \equiv -q^2
\label{Qsq}
\eeq
so that $Q^2 > 0$ in the spacelike region, and set 
\beq
\hat{q}_\mu \equiv \frac{q_\mu}{\sqrt{Q^2}}
\label{qhat}
\eeq
The bispinor basis elements can be chosen in such a manner
that they have the same spin, parity and charge conjugation as the
corresponding current $J_\mu^a(x)$.  For the vector vertex we adopt the
following bispinor basis elements:
\beqs
 &&  \Gamma^{(V)}_1 = \fsl{\epsilon},\ \ 
     \Gamma^{(V)}_2 = \frac{1}{2} [\fsl{\epsilon},\fsl{p}] 
                            (p \cdot \hat q), \ \ 
     \Gamma^{(V)}_3 = \frac{1}{2} [\fsl{\epsilon},\fsl{\hat q}], \cr\cr
 &&  \Gamma^{(V)}_4 = \frac{1}{3!}[\fsl{\epsilon},\fsl{p},\fsl{\hat q}],\ \ 
     \Gamma^{(V)}_5 = (\epsilon \cdot p),\ \ 
     \Gamma^{(V)}_6 = \fsl{p} (\epsilon \cdot p), \cr\cr
 &&  \Gamma^{(V)}_7 = \fsl{\hat q}(p \cdot \hat q) 
            (\epsilon \cdot p),\ \ 
     \Gamma^{(V)}_8 = \frac{1}{2} [\fsl{p},\fsl{\hat q}]
            (\epsilon \cdot p),
\label{vbispinors}
\eeqs
where $[a,b,c] \equiv a[b,c] + b[c,a] + c[a,b]$.  For the axial-vector vertex
we use the bispinor basis elements
\beqs
 && \hspace{-4mm}
     \Gamma^{(A)}_1 = \fsl{\epsilon}\ \gamma_5 ,\  
     \Gamma^{(A)}_2 = \frac{1}{2} [\fsl{\epsilon},\fsl{p}] 
                            \gamma_5  ,\  
     \Gamma^{(A)}_3 = \frac{1}{2} [\fsl{\epsilon},\fsl{\hat q}]\ 
              (p \cdot \hat q)
                 \ \gamma_5 , \cr\cr
 && \hspace{-4mm}
      \Gamma^{(A)}_4 = \frac{1}{3!}[\fsl{\epsilon},\fsl{p},
              \fsl{\hat q}] \gamma_5,\ \ 
     \Gamma^{(A)}_5 = (\epsilon \cdot p)\ (p \cdot \hat q)\ 
        \gamma_5, \cr\cr
 && \hspace{-4mm}
     \Gamma^{(A)}_6 = \fsl{p} (\epsilon \cdot p)\ \gamma_5,\ \ 
     \Gamma^{(A)}_7 = \fsl{\hat q}\ (\epsilon \cdot p)\ 
             (p \cdot \hat q)
                 \ \gamma_5, \cr\cr
 && \hspace{-4mm}
     \Gamma^{(A)}_8 = \frac{1}{2} [\fsl{p},\fsl{\hat q}]
             (\epsilon \cdot p)\
      (p \cdot \hat q)\ \gamma_5.
\label{avbspinors}
\eeqs
Given the charge conjugation properties of the vector and axial-vector
currents and the above choice of the bispinor basis elements, it 
follows that invariant amplitudes $\chi_i^{(J)}$ are even functions of $p\cdot
\hat{q}$.

In the present analysis it is convenient to choose the Lorentz reference frame
so that only the timelike component of $q^\mu$ is nonzero.  Since we are
working in the spacelike region (which avoids physical mass singularities), we
thus use a Wick rotation with 
\beq
   q^\mu = (i Q , 0 , 0 , 0 ) .
\label{qparam}
\eeq
Similarly, for the relative momentum $p^\mu$ we perform a Wick rotation and
parametrize it in terms of the real variables $u$ and $w$ (with
dimensions of mass) as
\beq
   p \cdot q = - Q\  u \ ,\ \  p^2 = - u^2 - w^2 .
\label{pparam}
\eeq
Hence, the invariant amplitudes $\chi^{(J)}_i$ are functions of $u$ and $w$:
\beq
   \chi_i^{(J)} = \chi_i^{(J)}(u,w;Q) .
\label{chidep}
\eeq
{}Owing to the charge-conjugation properties of the Bethe-Salpeter amplitude
$\chi^{(J)}$ and the bispinor basis elements defined above, the invariant
amplitudes $\chi_i^{(J)}(u,w)$ satisfies the relation
\beq
   \chi_i^{(J)}(u,w;Q) = \chi_i^{(J)}(-u,w;Q) \ .
 \label{eq:even-property}
\eeq
Using this property of the invariant amplitudes, we rewrite 
Eq.~(\ref{eq:Pi_JJ}) as 
\beqs
  && \hspace{-1.5cm}
   \Pi_{VV}(Q^2) \ =\  \frac{N}{\pi^3} \int_0^\infty du 
   \int_0^\infty dw\ w^2 \cr\cr
  &&\left[ - \chi^{(V)}_1(u,w;Q) + \frac{w^2}{3} \chi^{(V)}_6(u,w;Q) \ \right],
 \label{eq:PiVV}
 \eeqs
 \beqs
  && \hspace{-1.5cm}
   \Pi_{AA}(Q^2) \ =\  \frac{N}{\pi^3} \int_0^\infty du 
   \int_0^\infty dw\ w^2 \cr\cr
&&\left[\ \ \chi^{(A)}_1(u,w;Q)  - \frac{w^2}{3} \chi^{(A)}_6(u,w;Q) \ \right].
\label{eq:PiAA}
\eeqs
Here, we have used the expanded form of the Bethe-Salpeter amplitude given in 
Eq.~(\ref{eq:chi-expanded}) and carried out the three-dimensional 
angular integration.
 
 From Eqs.~(\ref{eq:PiVV}) and (\ref{eq:PiAA}), it follows that 
\beqs
 \Pi_{V-A}(Q^2) &=&
 \frac{1}{3}\left (\frac{N}{2}\right )\sum_{\epsilon} \int \frac{d^4 p}{i (2
 \pi)^4}  \cr\cr
 &&{\rm Tr} \bigg [\fsl{\epsilon} \chi^{(J)}(p;q,\epsilon) -
 \fsl{\epsilon} \gamma_5 \chi^{(A)}(p;q,\epsilon) \bigg ] \cr\cr
 &=&  \frac{N}{\pi^3} \int_0^\infty du \int_0^\infty dw\ w^2 \cr\cr 
 &&\Bigg[ - \bigg( \chi^{(V)}_1(u,w;Q) + \chi^{(A)}_1(u,w;Q) \bigg ) \cr\cr
 && + \frac{w^2}{3} \bigg ( \chi^{(V)}_6(u,w;Q) + \chi^{(A)}_6(u,w;Q) \bigg )
 \ \Bigg]. \cr\cr
& & 
\label{eq:PiV-A}
\eeqs 
Although both $\Pi_{VV}(Q^2)$ and $\Pi_{AA}(Q^2)$ are individually
logarithmically divergent, the underlying ${\rm SU}(N_f)_L \times {\rm
SU}(N_f)_R$ chiral symmetry guarantees that these divergences cancel in the
difference $\Pi_{VV} - \Pi_{AA}$, which is therefore finite.

\section{Inhomogeneous Bethe-Salpeter equation}
\label{sec:IBS}
 
\begin{figure}[t]
   \begin{center}
     \includegraphics[height=1.8cm]{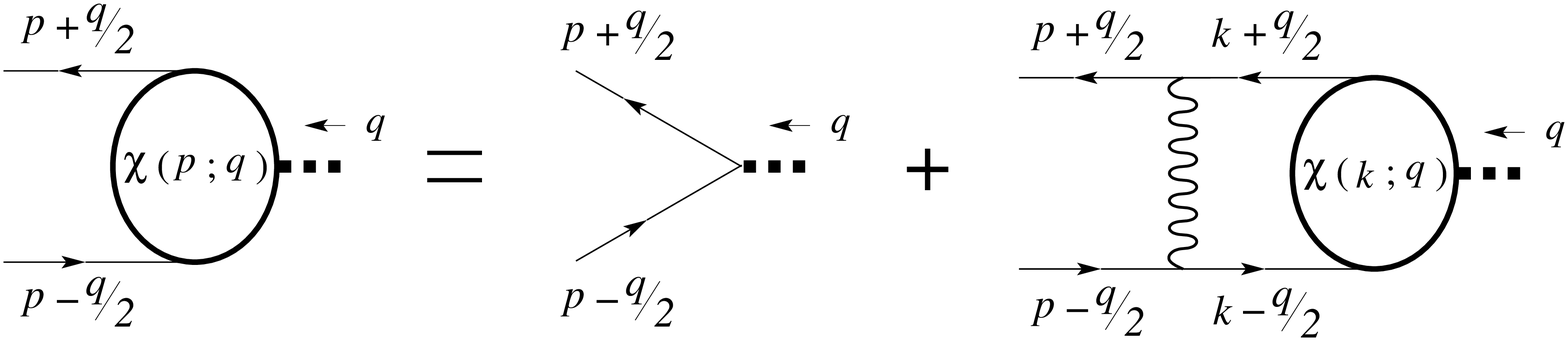}
   \end{center}
 \caption{A graphical expression of the IBS equation in the 
 (improved) ladder approximation.}
 \label{fig:IBSeq}
 \end{figure}

In this section we discuss the full (inhomogeneous) Bethe-Salpeter equation,
which we will use to calculate current-current correlation functions and, from
these, $S$ (and, as a check, also $f_P$).  The IBS equation is a
self-consistent description of the coupling of a current $J^a_\mu$ to
fermion-antifermion bound states.  This coupling is represented by the
Bethe-Salpeter amplitude $\chi^{(J)}$.  The four-momenta assigned to the
fermion and antifermion are $p_\psi = p + (q/2)$ and $p_{\bar \psi}=-p+(q/2)$
so that the total momentum of the bound state is $q$, and the relative momentum
of the $\psi$ and $\bar \psi$ is $2p$. Since we are dealing with $J=1$ bound
states, the bound-state amplitude also depends on the polarization vector
$\epsilon$, which satisfies $\epsilon \cdot q = 0$ and $\epsilon \cdot \epsilon
= -1$.  A grapical indication of the inhomogeneous Bethe-Salpeter equation
structure is given in Fig.~\ref{fig:IBSeq}. 
The IBS equation is 
\beq
   T(p;q)\ \chi^{(J)}(p;q,\epsilon) \ \ =\ \ 
   \epsilon \cdot G^{(J)}\ 
   +\ K(p;k) \ast \chi^{(J)}(k;q,\epsilon) .
 \label{eq:IBSeq}
\eeq
Here the kinetic part $T$ is
\beq
  T(p;q) =   - S_F^{-1}(p + q/2) \otimes  S_F^{-1}(p - q/2) \ ,
 \label{Tdef}
\eeq
where $S_F(p)=1/(A(p)\fsl{p} - \Sigma(p))$. We follow the standard procedure of
using the Landau gauge in calculations with the Schwinger-Dyson and
Bethe-Salpeter equations since in this gauge the fermion wave function
renormalization factor $A(p)=1$.  The physical results are, of course,
gauge-invariant. The Bethe-Salpeter kernel $K$ in the improved ladder
approximation is expressed as
\beqs
\hspace{-1cm}
 K(p;k) &=&  C_{2f} \, \frac{\bar{g}^2(p,k)}{(p-k)^{2}} \times \cr\cr 
  &\times&  \left(-g_{\mu\nu} + \frac{(p-k)_\mu (p-k)_\nu}{(p-k)^2}
          \right) \cdot \gamma^\mu \otimes \gamma^\nu ,
\label{Kdef}
\eeqs
where $C_{2f}$ was given in Eq. (\ref{c2f}). 
In the above expressions we use the tensor product notation
\beq
   (A \otimes B) \,\chi  =  A\, \chi\, B \ ,
\label{aob}
\eeq
and the inner product notation
\beq
  K(p;k) \ast \chi^{(J)}(k;q,\epsilon) =   
   \int \frac{d^4 k}{i(2\pi)^4}\  K(p,k)\  \chi(k;q) \ .
\label{kchi}
\eeq
where summations over Dirac indices are understood.  (In contrast, for our
previous calculations \cite{mmw,sg} of meson masses, we only needed to use the
homogeneous Bethe-Salpeter equation.)

As in Refs. \cite{Kugo:1992pr,mmw,pms,pmsw,sg}, we make the ansatz
for the running coupling, after Euclidean rotation,
\beq
\alpha(p_E,q_E) = \alpha(p_E^2+q_E^2) \ ,
\label{gsqform}
\eeq
where the subscript denotes Euclidean.  Since $\alpha$ would naturally depend
on the gluon momentum squared, $(p-q)^2 = p^2+q^2-2p \cdot q$, the functional
form (\ref{gsqform}) amounts to dropping the scalar product term, $-2p \cdot
q$.  As discussed in Ref. \cite{sg}, this is a particularly reasonable
approximation in the case of a walking gauge theory because most of the
contribution to the integral comes from a region of Euclidean
momenta where $\alpha$ is nearly constant.  Hence, the shift upward or downward
due to the $-2p \cdot q$ term in the argument of $\alpha$ has very little
effect on the value of this coupling for the range of momenta that make the
most important contribution to the integral. The approximation (\ref{gsqform})
enables one to carry out the angular integration analytically. 

The momentum-dependent dynamical mass $\Sigma(p)$ for the fermion is obtained
from the Schwinger-Dyson equation,
\beq
   \Sigma(p) = - K(p,k) \ast S_F(p).
\label{sdeq}
\eeq
We use the same kernel $K(p,k)$ here as in the IBS equation in order to respect
the ${\rm SU}(N_f)_L \times {\rm SU}(N_f)_R$ chiral symmetry~\cite{
Kugo:1992pr,Bando-Harada-Kugo}.  The numerical method that is used for solving
the SD equation and the IBS equation involves approximating the respective
integrals by discrete sums and is the same as in Ref.~\cite{pms,pmsw}.
The reader is referred to these papers for more details on this method.

\section{Results and Discussion}
\label{sec:results}

In this section we present the results of our calculations of $\Pi_{V-A}(Q^2)$
and $\hat S$ using the Schwinger-Dyson and (inhomogeneous) Bethe-Salpeter
equations.  It is appropriate to include an obvious cautionary remark that
these calculations involve strong couplings $\alpha$ of order unity, and
therefore there could be significant corrections to the (improved) ladder
approximation used in our solutions of the Schwinger-Dyson and Bethe-Salpeter
equations.  Accordingly, as one check on the reliability of our methods, we
have also carried out a comparison of $f_P$ obtained from the SD and IBS
equations via eq. (\ref{w1}) with $f_P$ obtained from the SD equation via the
Pagels-Stokar relation.  

\begin{figure}[t]
   \begin{center}
     \includegraphics[height=6.2cm]{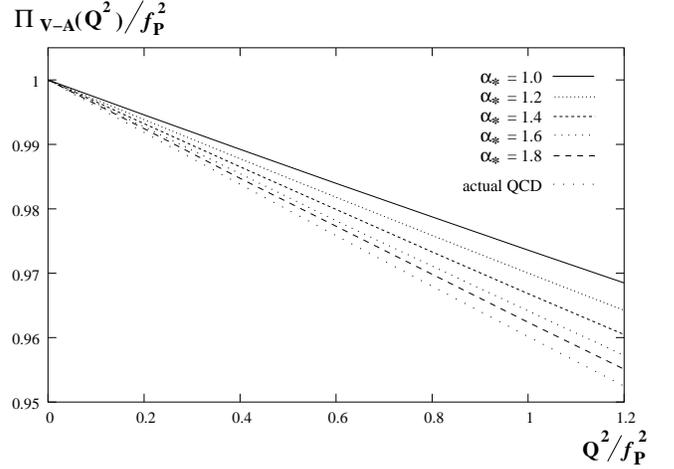}
   \end{center}
 \caption{Plot of $\Pi_{V-A}(Q^2)/f_P^2$ as a function of $Q^2$ in the
spacelike interval $0 \le Q^2/f_P^2 \le 1.2$, for $\alpha_* = 1.0, 1.2, 1.4,
1.6, 1.8$.  As indicated, the horizontal axis refers to the quantity
$Q^2/f_P^2$.  For comparison, $\Pi_{V-A}(Q^2)/f_P^2$ for QCD with $N_f=3$ 
massless quarks is also plotted.  See text for further details.}
 \label{fig:PiV-A}
 \end{figure}

In Fig.\ref{fig:PiV-A} we plot our calculated values of $\Pi_{V-A}(Q^2)/f_P^2$
for $\alpha_* = 1.8, 1.6, 1.4, 1.2$ and 1.0, as functions of the dimensionless
quantity $Q^2/f_P^2$.
The slope of each curve at $Q^2=0$ is equal to $-{\hat S}/(4\pi)$ for the given
value of $\alpha_*$.  We first observe that $\Pi_{V-A}(Q^2)$ is almost linear
at $Q^2=0$, with a small positive second derivative.  This is the justification
for our earlier statement that the implications of our findings for technicolor
would be essentially the same whether we used the definition of $\hat S$ in
terms of a derivative at $Q^2=0$ or a finite difference, analogously to
Eq. (\ref{spdg}). Secondly, from Fig.  \ref{fig:PiV-A} it is clear that the
magnitude of the slope, and hence $\hat S$, decreases as $\alpha_*$ decreases
toward the chiral phase transition point $\alpha_{cr}$.

\begin{figure}[t]
   \begin{center}
     \includegraphics[height=6.4cm]{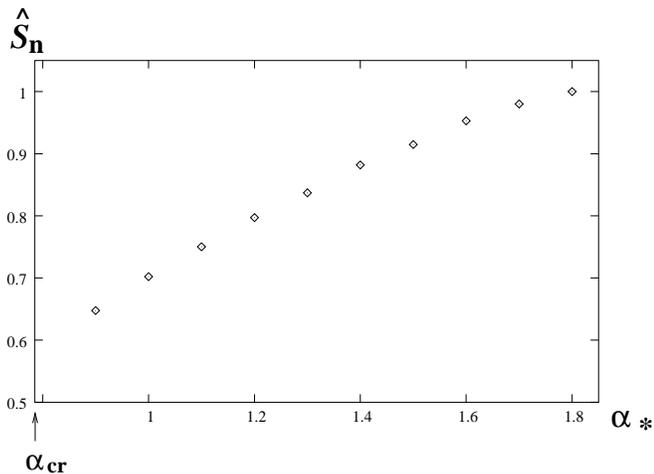}
   \end{center}
 \caption{Plot of ${\hat S}_n$ for several values of $\alpha_*$ in the 
range $0.9 \le \alpha_\ast \le 1.8$. As indicated by the subscript $n$, the 
values are normalized by the value of $\hat S$ at $\alpha_*=1.8$.}
 \label{fig:S-hat_n}
\end{figure}

In Fig.~\ref{fig:S-hat_n}, we plot the values of ${\hat S}$, derived from the
slope of $\Pi_{V-A}(Q^2)$ at $Q^2=0$, for several values of $\alpha_*$.  As
indicated by the subscript $n$, the values are normalized by the value of $\hat
S$ at $\alpha_*=1.8$.
This figure shows that $\hat S$ decreases by about 40 \% as $\alpha_*$ is
reduced from 1.8 to 0.9, or equivalently (c.f. Eq. (\ref{nfsol})) as $N_f$ is
increased from 10.3 to 11.6.  As our calculation of meson masses in
Ref. \cite{sg} showed, this is a crossover region, in which the theory is
changing from QCD-like, non-walking behavior at smaller $N_f$ to the walking
regime at larger $N_f$ approaching $N_{f,cr}$.  Reinserting the factor of
$N_D=N_f/2$ to get $S$ itself, we obtain a decrease by about 30 \% in $S$,
since $N_D$ only increases by about 10 \% over this range.  Thus, our
calculation shows that for this range of values, $\hat S$ decreases
significantly as one moves from the QCD-like to the walking regimes.  This
finding is an important result of our present study.

\begin{figure}[t]
   \begin{center}
     \includegraphics[height=6.4cm]{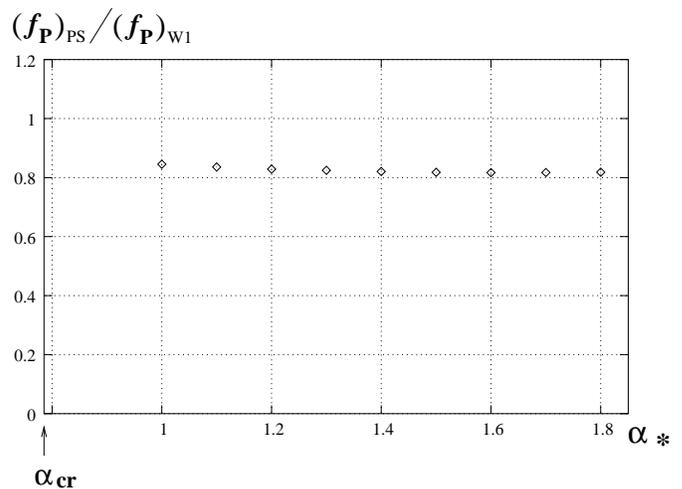}
   \end{center}
 \caption{Plot of the ratio $(f_P)_{PS}/(f_P)_{W1}$, where $(f_P)_{PS}$ and 
$(f_P)_{W1}$ denote $f_P$ calculated via Eq. (\ref{psrel}) and via 
Eq. (\ref{w1}), respectively, as a function of $\alpha_*$.}
 \label{fig:fpi_ratio}
\end{figure}

As a check on our calculational methods, we compare $f_P$ calculated in two
different ways: via the Pagels-Stokar relation, Eq. (\ref{psrel}), denoted
$(f_P)_{PS}$, and by the first Weinberg sum rule (W1) or equivalent relation
$\Pi_{V-A}(0)=f_P^2$ in Eq. (\ref{w1}), denoted $(f_P)_{W1}$.  In view of the
fact that the Pagels-Stokar itself is approximate and that our solution of the
SD and IBS equations involves the ladder approximation and the neglect of
completely nonperturbative contributions such as those due to instantons, we do
not expect exact agreement between these two different methods of calculation.
In Fig. \ref{fig:fpi_ratio} we present a plot of the ratio
$(f_P)_{PS}/(f_P)_{W1}$ as a function of $\alpha_*$.  The closeness of this
ratio to unity gives one measure of the accuracy and reliability of our
calculations.
We see that the ratio is within about 20 \% of unity and is essentially
independent of $\alpha_*$ in the range considered, with the Pagels-Stokar
method yielding a slightly smaller value than the expression in terms of the
current-current correlation functions.  This gives us further confidence in the
results of our SD-IBS calculation of $\hat S$.

Our finding that $\hat S$ and $S$ are reduced as one moves from the QCD-like
regime toward the walking regime of an SU($N$) gauge theory is in agreement
with the approximate analytic results of Refs. \cite{ats,acd,as_s,iwts}, and 
it complements those works, being based on a numerical solution of the
Schwinger-Dyson and Bethe-Salpeter equations. In the sub-interval 
$0.9 \le \alpha_* \le 1.0$ closer to the walking limit our results coincide 
with those in Ref. \cite{pmsw}.  Our use of a larger interval
has the advantage that we are able to observe a larger reduction in $\hat S$ as
$\alpha_*$ decreases than was done in Ref.  \cite{pmsw}.  We have not attempted
here to examine the extreme walking limit $(\alpha_* - \alpha_{cr})/\alpha_{cr}
\to 0^+$.  As discussed in Ref. \cite{pmsw}, it becomes increasingly difficult
to obtain an accurate numerical solution for $\hat S$ in this limit because of
very strong cancellations.

In addition to this decreasing trend of $\hat S$, one may also discuss the
absolute magnitude of $\hat S$.  Our calculation yields $\hat S = 0.47$ at
$\alpha_*=1.8$. If $\hat S$ continues to be a monotonically function of $N_f$
(and hence, in the range of interest here, also a monotonic function of
$\alpha_*$), then an extrapolation of our calculation to the QCD case with
$N_f=2$ or, with the strange quark mass neglected, $N_f=3$, would predict a
value of $\hat S \gsim 0.5$.  This is similar to the value that was obtained in
earlier studies using the SD and IBS equations in a different manner than here,
where it was necessary to introduce an cutoff on the growth of the strong
coupling in the infrared \cite{hys,pms}.  In Ref. \cite{pms} it was shown that
if one used a cutoff that led to a very large value of the coupling, one could
get a result for $S$ in agreement with the experimental value (\ref{svalue}),
but the reliability of the calculational method in the presence of such a large
coupling was not clear.  We shall adopt the optimisitc viewpoint here of giving
greater weight to the change in $\hat S$ as a function of $N_f$ than to the
absolute value of $\hat S$ itself.  Equivalently, one could envision applying
an overall correction factor of about 2/3 to the absolute magnitude of $\hat S$
so that the value for small $N_f$ matches that in QCD.  Physically, this
factor would be regarded as correcting for the strong-coupling effects not
included in the SD-IBS analysis.

Although one cannot use perturbation theory reliably to calculate $S$ in a
strongly coupled gauge theory, the perturbative result is often used in
discussions of constraints on new physics, and hence it is worthwhile to see
how our results compare with the perturbative computation. A one-loop
perturbative calculation with degenerate fermions having effective masses
satisfying $(2\Sigma/m_Z)^2 \gg 1$ yields the well-known result
$S_{pert.}=N_{D,tot.}/(6 \pi)$ where here $N_{D,tot.}=N_D N$, i.e,
\beq
\hat S_{pert.} = \frac{N}{6\pi} \ . 
\label{s_hat_pert}
\eeq
In QCD with $N_f=2$ and $N=N_c=3$, this perturbative calculation would predict
$S_{QCD,pert.} \simeq 1/(2\pi) \simeq 0.16$.  The experimental value in
Eq. (\ref{svalue}) is approximately twice this; $S_{QCD} \simeq 2
S_{QCD,pert.}$.  The reductions that we have found in $\hat S$ and $S$ for the
range of $\alpha_*$ investigated suggest that in a walking theory, much or all
of the above factor of 2 might be removed, and the true value of $S$ might well
be comparable to, or, indeed, perhaps less than, the perturbative estimate.

For reference, in the one-doublet and one-family technicolor models, the
perturbative expressions for the techniparticle contributions to $S$ are
$S_{TC,pert.}=N_{TC}/(6\pi)$ and $S_{TC,pert.}=2N_{TC}/(3\pi)$, respectively.
With $N_{TC}=2$, these take the values $S_{QCD,pert.} \simeq 0.1$ and 0.4,
respectively. Fits to precision electroweak data yield allowed regions in $S$
and two other parameters describing modifications of the $Z$ and $W$
propagators by new physics beyond the Standard Model, namely the parameter $T$
measuring violations of custodial SU(2) from this new physics and a third
parameter, $U$, of somewhat less importance here.  Since the Standard Model
expression for $S$ includes a term $(1/(6\pi))\ln (m_H/m_{H,ref.})$, the
resultant allowed regions depend on the choice of the reference value of the SM
Higgs mass, $m_{H,ref.}$.  The comparison of these with a technicolor theory is
complicated by the fact that technicolor has no fundamental Higgs field;
sometimes one formally uses $m_{H,ref.} \sim 1$ TeV for a rough estimate, since
the SM with $m_H \sim 1$ TeV has strong longitudinal vector boson scattering,
as does technicolor. However, this may involve some double-counting when one
also includes contributions to $S$ from technifermions, whose interactions and
bound states (e.g., techni-vector mesons) are responsible for the strong $W^+_L
W^-_L$ and $Z_LZ_L$ scattering in a technicolor framework.  The current fit
\cite{pdg,lepewwg} disfavors values of $S \gsim 0.2$.  Our findings in this
paper suggest, in agreement with the previous works noted above using different
methods, that the constraint on walking technicolor models could be less severe
than would be inferred from the perturbative formula for the technifermion
contribution to $S$.

\section{Summary}
\label{sec:Summary}

In summary, using numerical solutions of the Schwinger-Dyson and
(inhomogeneous) Bethe-Salpeter equation, we have calculated the $S$ parameter
as a function of the approximate infrared fixed point, $\alpha_*$, or
equivalently, the number of massless fermions, $N_f$, in a vectorial, confining
SU($N$) gauge theory.  We have focused on the crossover region between the
walking and QCD-like (non-walking) regimes.  Our results show that $\hat S$ and
also $S$ decrease significantly as $\alpha_*$ decreases in this range.  This
trend agrees with earlier indications of a decrease in $S$ in walking gauge
theories.  We have discussed the implications for technicolor models.

\bigskip

M.K. thanks Profs. M. Harada and K. Yamawaki for the collaborations on the
related Refs. \cite{mmw,pmsw}. R.S. thanks T. Appelquist for useful comments.  
This research was partially supported by the grant NSF-PHY-03-54776.

\section{Appendix}
\label{appendix}

In this appendix we present some results on vector meson dominance (VMD) and
their relevance to the crossover region that we are studying numerically.  In
QCD itself, VMD has served as a useful approximate model.  The matrix elements
for the production of a vector or axial-vector meson from the vacuum are 
\beq
\langle \rho^a | V^b_\mu | 0 \rangle = i \delta^{ab} f_\rho \epsilon_\mu
\label{rhomatrixelement}
\eeq
\beq
\langle a_1^a | A^b_\mu | 0 \rangle = i \delta^{ab} f_{a_1} \epsilon_\mu
\label{a1matrixelement}
\eeq
where $\epsilon_\mu$ are the respective polarization four-vectors. In QCD,
$f_\rho \simeq 154$ MeV, so that the first Weinberg sum rule yields 
$f_{a_1} =\sqrt{f_\rho^2-f_\pi^2} = 123$ MeV. A pole-dominated form for
the current-current correlation functions in the (simplistic) narrow-resonance
approximation is
\beq
\Pi_{V-A}(q^2) = \frac{f_V^2 m_V^2}{m_V^2-q^2-i\epsilon} 
               - \frac{f_A^2 m_A^2}{m_A^2-q^2-i\epsilon} \ , 
\label{piform}
\eeq
where, for QCD, the ground-state vector and axial-vector mesons are the
$\rho(776)$ and $a_1(1230)$, i.e., $V = \rho$ and $A = a_1$.  
Using the principal value relation
\beq
\frac{1}{s-s_0 \mp i\epsilon} = P \bigg ( \frac{1}{s-s_0} \bigg ) 
\pm i \pi \delta(s-s_0) \ , 
\label{pvrel}
\eeq
the pole-dominated form (\ref{piform}) gives, for the spectral functions, 
\beq
\rho_V(s) = f_V^2 \delta(s-m_V^2)
\label{rho_v}
\eeq
and
\beq
\rho_A(s) = f_A^2 \delta(s-m_A^2)
\label{rho_a}
\eeq
Although these are useful simplifications, more realistic analyses of the
spectral function sum rules take into account the finite widths of the $\rho$
and $a_1$ resonances, and also nonresonant contributions \cite{dg}. 

Expanding Eq. (\ref{piform}) for large Euclidean $q^2$, one obtains
\beq
\Pi_{V-A}(q^2) = -\frac{(f_V^2 m_V^2 - f_A^2 m_A^2)}{q^2} + 
O \Bigg ( \frac{1}{q^4} \Bigg ) \ . 
\label{piexpand}
\eeq
The asymptotic freedom of the SU($N$) gauge theory implies that for $|q^2| \gg
\Lambda^2$, $\Pi_{V-A}(q^2) \propto 1/q^4$, so this yields  the relation
\beq
f_V^2 m_V^2 - f_A^2 m_A^2 = 0 \ , 
\label{w2vmd}
\eeq
which is the second Weinberg sum rule, evaluated in the VMD approximation. 
Substituting the forms for the spectral functions into Eq. (\ref{w0}) yields 
\beq
S_{VMD} = -16\pi \bar L_{10} = 4\pi \Bigg ( \frac{f_V^2}{m_V^2}
                                           -\frac{f_A^2}{m_A^2} \Bigg ) 
\label{svmd}
\eeq
Similar substitutions into the first and second Weinberg sum rules yield the
relations 
\beq
f_V^2-f_A^2=f_P^2 
\label{w1vmd}
\eeq
and Eq. (\ref{w2vmd}).  The narrow-width VMD relations are approximate, since
they neglect the sizeable widths of the $\rho$ and $a_1$, but the Weinberg
relations follow from the asymptotic freedom of QCD when one sets $m_u=m_d=0$.

As $N_f$ increases and there is a crossover in the SU($N$) theory from QCD-like
behavior to walking behavior, there are resultant changes in how these integral
relations are satisfied.  For both the QCD-like and walking regimes, the
underlying asymptotic freedom of the theory implies that the current-current
correlation functions $\Pi_{V-A}(q^2)$ have a $1/q^4$ asymptotic falloff (up to
logs) for Euclidean $q^2 \gg \Lambda^2$.  However, in a walking regime, the
scale of chiral symmetry breaking, given by $f_P \sim \Sigma$, is much less
than $\Lambda$, and in the interval $\Sigma^2 \ll q^2 \ll \Lambda^2$,
$\Pi_{V-A}(q^2)$ have a less rapid falloff, $\sim 1/q^{4-2\gamma})$, where
$\gamma$ is the anomalous dimension of the bilinear fermion operator and is
$\gamma \simeq 1$ for strong walking, whence a $\Pi_{JJ}(q^2) \sim 1/q^2$
falloff in this interval. Physically, the coupling $\alpha$ is strong, of order
O(1) but slowly varying for an extended interval of energies between $\Sigma$
and $\Lambda$.  Thus it is plausible that in a walking theory, rather than just
the lowest-lying vector and axial-vector resonances making important
contributions to the various spectral function integrals, there could be
significant contributions from a number of higher-lying states with the same
quantum numbers \cite{lanerev}.  In this case, still retaining the VDM
approximation, one would generalize the expressions above to be sums over these
higher-lying states.  For example, the evaluation of Eqs. (\ref{w0}),
(\ref{w1}), and (\ref{w2}) in the VMD approximation would read
\beq
\hat S_{VMD} =  4\pi \sum_i \Bigg ( \frac{f_{V_i} ^2}{m_{V_i}^2}
                                   -\frac{f_{A_i}^2}{m_{A_i}^2} \Bigg ) 
\label{svmdwalk}
\eeq
\beq
\sum_i(f_{V_i}^2-f_{A_i}^2)=f_P^2 
\label{w1vmdwalk}
\eeq
and
\beq
\sum_i (f_{V_i}^2 m_{V_i}^2 - f_{A_i}^2 m_{A_i}^2) = 0 \ . 
\label{w2vmdwalk}
\eeq
More generally, however, it is not clear that the VMD model would apply in this
walking regime.  One analytic approximation is to use VMD for a partial
evaluation representing contributions from $\sqrt{s} \simeq \Sigma$ and
supplement this with a term due to a fermion loop \cite{as_s}.  In the region
that we have concentrated on in this paper, in which the theory is crossing
over between strong walking behavior and QCD-like behavior, one expects that
the saturation of the spectral function integrals has a form that is
intermediate between walking and QCD-like.  In our actual calculation of $\hat
S$ via Eq. (\ref{scor}), we only need to calculate $\Pi_{V-A}'(0)$, which we do
by means of numerical solutions of the relevant SD and IBS equations, so we do
not have to deal with questions of how the spectral functions behave in a
walking gauge theory. In future work it would be worthwhile to use the
connection between these two different ways of calculating $S$ to undersand the
hadronic spectrum better in such a walking gauge theory.


\begin{thebibliography}{99}

\bibitem{bz}
An early paper on the phase structure of vectorial gauge theories is 
T.~Banks and A.~Zaks,
Nucl.\ Phys.\ B {\bf 196} (1982), 189.

\bibitem{tc}
S. Weinberg, Phys. Rev. D {\bf 19}, 1277 (1979);
L. Susskind, {\it ibid.} D {\bf 20}, 2619 (1979); see also 
S. Weinberg, Phys. Rev. D {\bf 13}, 974 (1976).

\bibitem{wtc1}
B. Holdom, Phys. Lett. B {\bf 150}, 301 (1985).

\bibitem{wtc2}
K. Yamawaki, M. Bando, and K. Matumoto, Phys. Rev. Lett. {\bf
56}, 1335 (1986).

\bibitem{chipt1}
T. Appelquist, D. Karabali, and L. C. R. Wijewardhana, Phys. Rev. Lett. {\bf
57}, 957 (1986); T. Appelquist and L. C. R. Wijewardhana, Phys. Rev. D
{\bf 35}, 774 (1987); Phys. Rev. D {\bf 36}, 568 (1987).

\bibitem{chipt2}
T. Appelquist, J. Terning, and L. C. R. Wijewardhana,
Phys. Rev. Lett.  {\bf 77}, 1214 (1996).

\bibitem{my}
V. Miransky and K. Yamawaki, Phys. Rev. D {\bf 55}, 5051 (1997); {\it ibid.}
{\bf 56}, E 3768 (1997). See also V. Miransky and P. Fomin,
Sov. J. Part. Nucl. {\bf 16}, 203 (1985).

\bibitem{chipt3}
T. Appelquist, A. Ratnaweera, J. Terning, and
L. C. R. Wijewardhana, Phys. Rev. D {\bf 58}, 105017 (1998).

\bibitem{gl85}
J. Gasser and H. Leutwyler, Nucl. Phys. B {\bf 250}, 465 (1985); 
{\it ibid.} B {\bf 250}, 517 (1985). See also the papers in
Refs. \cite{chipt}-\cite{hyrev}. 

\bibitem{chipt}
J. Gasser and H. Leutwyler, Phys. Rept. C {\bf 87}, 77 (1982);
J. Gasser, H. Leutwyler, A. Pich, and E. de Rafael, Nucl. Phys. {\bf 321}, 311
(1989); G. Colangelo, J. Gasser, and H. Leutwyler, Nucl. Phys. B {\bf 603}, 
125 (2001). 

\bibitem{resonances}
G. Ecker, J. Gasser, A. Pich, and E. de Rafael, Nucl. Phys. B {\bf 321}, 311
(1989); G. Ecker, J. Gasser, H. Leutwyler, A. Pich, and E. de Rafael,
Phys. Lett. B {\bf 223}, 425 (1989); M. Knecht and E. de Rafael, 
Phys. Lett. B {\bf 424}, 335 (1998). 

\bibitem{meissner}
U. Meissner, Rept. Prog. Phys. {\bf 56}, 903 (1993). 

\bibitem{hyrev}
M. Harada and K. Yamawaki, Phys. Rept. {\bf 381}, 1 (2004).

\bibitem{pt}
M.~E.~Peskin and T.~Takeuchi,
Phys.\ Rev.\ Lett.\  {\bf 65}, 964 (1990);
Phys.\ Rev.\ D {\bf 46}, 381 (1992). 

\bibitem{ab}
G. Altarelli and R. Barbieri, Phys. Lett. B {\bf 253}, 161 (1991);
G. Altarelli, R. Barbieri, F. Caravaglios,
Int. J. Mod. Phys. A {\bf 13}, 1031 (1998).

\bibitem{pdg}
See http://pdg.lbl.gov. 

\bibitem{lepewwg}
http://lepewwg.web.cern.ch/LEPEWWG/plots.

\bibitem{scalc1}
B. Holdom and J. Terning, Phys. Lett. B {\bf 247}, 88 (1990); 
M. Golden and L. Randall, Nucl. Phys. {\bf B} {\bf 361}, 3 (1991);
R. Johnson, B.-L. Young, and D. McKay, Phys. Rev. D {\bf 43} (1991) R17;
R. Cahn and M. Suzuki, Phys. Rev. D {\bf 44}, 3641 (1991).

\bibitem{ats}
T. Appelquist and G. Triantaphyllou, Phys. Lett. B {\bf 278}, 345 (1992).

\bibitem{acd}
R. Sundrum and S. Hsu, Nucl. Phys. B {\bf 391}, 127 (1993).

\bibitem{hys}
M.~Harada and Y.~Yoshida,
Phys.\ Rev.\ D {\bf 50}, 6902 (1994). 

\bibitem{as_s}
T. Appelquist and F. Sannino, Phys. Rev. D {\bf 59}, 067702 (1999).

\bibitem{iwts}
S. Ignjatovic, L. C. R. Wijewardhana, and T. Takeuchi, Phys. Rev. 
D {\bf 61}, 056006 (2000).

\bibitem{pmsw}
M. Harada, M. Kurachi, and K. Yamawaki, Prog. Theor. Phys. {\bf 115}, 765
(2006); see also {\it  Proc. of 2004 International Workshop on Dynamical 
Symmetry Breaking, 2004}, eds. M. Harada and K. Yamawaki, 
(Nagoya Univ., 2005), p. 125.

\bibitem{sg}
M.~Kurachi and R.~Shrock,
hep-ph/0605290.

\bibitem{mmw} 
M.~Harada, M.~Kurachi and K.~Yamawaki,
Phys.\ Rev.\ D {\bf 68}, 076001 (2003).

\bibitem{mme}
K.~I.~Aoki, M.~Bando, T.~Kugo, M.~G.~Mitchard and H.~Nakatani,
Prog. Theor. Phys. {bf 84}, 683 (1990); T.~Kugo, {\it in Proc. of 1991 Nagoya
Spring School on Dynamical Symmetry Breaking}, ed. K. Yamawaki (World
Scientific Pub. Co., Singapore, 1992), p. 35; K.~I.~Aoki, T.~Kugo and
M.~G.~Mitchard, Phys. Lett. B {\bf 266}, 467 (1991); Phys. Lett. B {bf 286}, 
355 (1992).

\bibitem{pms}
M.~Harada, M.~Kurachi and K.~Yamawaki,
Phys.\ Rev.\ D {\bf 70}, 033009 (2004).

\bibitem{integer}
%
Here and below, when we mention non-integral values of $N_f$, it is implicitly
understood that physical values of $N_f$ are, of course, non-negative integers,
and the non-integral values are defined via an analytic continuation away from
these physical values.

\bibitem{b0}
D. Gross and F. Wilczek, Phys. Rev. Lett. {\bf 30}, 1343 (1973);
H. D. Politzer, Phys. Rev. lett. {\bf 30}, 1346 (1973). 

\bibitem{b1}
W. Caswell, Phys. Rev. Lett. {\bf 33}, 244 (1974); D. R. T. Jones, Nucl. Phys.
B {\bf 75}, 531 (1974).

\bibitem{casimir} The Casimir invariant $C_2(R)$ of the representation $R$ is
defined by ${\cal D}_R(T_a)^i_j{\cal D}_R(T_a)^j_k = C_2(R)\delta^i_k$, where
$\{a,b\}$ and $\{i,j,k\}$ denote group and representation indices and sums over
repeated indices are understood.

\bibitem{alm}
T. Appelquist, K. Lane, and U. Mahanta, Phys. Rev. Lett. {\bf
61}, 1553 (1988); T. Appelquist, U. Mahanta, D. Nash, and L.C.R.  Wijewardhana,
Phys. Rev. D {\bf 43}, 646 (1991);
U. Mahanta, Phys. Rev. D {\bf 45}, 1405 (1992).

\bibitem{lgt} 
Y.  Iwasaki et al., Phys. Rev. Lett. {\bf 69}, 21 (1992);
Phys. Rev. D {\bf 69}, 014507 (2004); 
P. Damgaard, U. Heller, A. Krasnitz, and P. Olesen, Phys. Lett. B {\bf 400},
169 (1997); R.  Mawhinney, Nucl. Phys. B (Proc. Suppl.) {\bf 63A-C}, 212 
(1998); R.  Mawhinney, Nucl. Phys. B (Proc. Suppl.) {\bf 83}, 57 (2000).

\bibitem{Gardi}
E.~Gardi and M.~Karliner,
E.~Gardi, G.~Grunberg and M.~Karliner,

\bibitem{ExplicitSolution}
R.~M.~Corless, G.~H.~Gonnet, D.~G.~E.~Hare, D.~J.~Jeffrey and D.~E.~Knuth,
Adv.\ Comput.\ Math.\  {\bf 5}, 329 (1996).

\bibitem{etc}
S. Dimopoulos, L. Susskind, Nucl. Phys. {\bf B155}, 23, (1979);
E. Eichten, K. Lane, Phys.  Lett. B {\bf 90}, 125 (1980);
E.~Farhi and L.~Susskind, Phys. Rep. {\bf 74}, 277 (1981). 

\bibitem{tcm} 
Some recent reviews of TC/ETC models are given in
\cite{lanerev,etcrev} and some recent studies include \cite{at94,as,sann,lm}.
We do not consider here models that hypothesize a strong (``topcolor'') 
interaction that produces a $\langle \bar t t \rangle$ condensate; these are
reviewed in Refs. \cite{lanerev,etcrev}. 

\bibitem{lanerev}
K. Lane, hep-ph/0202255.

\bibitem{etcrev}
C. Hill and E. Simmons, Phys. Rep. {\bf 381}, 235 (2003); 
R. S. Chivukula, M. Narain, and J. Womersley, in Ref. \cite{pdg}.

\bibitem{at94}
T. Appelquist and J. Terning, Phys. Rev. D {\bf 50}, 2116 (1994).

\bibitem{as}
T. Appelquist and R. Shrock, Phys. Lett. B {\bf 548}, 204 (2002); 
Phys. Rev. Lett. {\bf 90}, 201801 (2003); T. Appelquist, M. Piai, and
R. Shrock, Phys. Rev. D {\bf 69}, 015002 (2004); {\it ibid.}  
D {\bf 70}, 093010 (2004);  Phys. Lett. B {\bf 593}, 175 (2004); 
{\it ibid.} B {\bf 595}, 442 (2004). 

\bibitem{sann}
D. Hong, S. Hsu, and F. Sannino, Phys. Lett. B {\bf 597}, 89 (2004); 
D. Dietrich, F. Sannino, K. Tuominen, Phys. Rev. D {\bf 72}, 055001 (2005). 

\bibitem{lm}
K. Lane and A. Martin, Phys. Rev. D {\bf 71}, 076007 (2005); 
Phys. Lett. B {\bf 635}, 118 (2006). 

\bibitem{ts}
N. D. Christensen and R. Shrock, Phys. Lett. B {\bf 632}, 92 (2006); see also
N. D. Christensen and R. Shrock, Phys. Rev. Lett. {\bf 94}, 241801 (2005).

\bibitem{dmo}
T.~Das, V.~S.~Mathur and S.~Okubo,
Phys.\ Rev.\ Lett.\  {\bf 19}, 859 (1967).

\bibitem{psrel}
H. Pagels and S. Stokar, Phys. Rev. D {\bf 20}, 2947 (1979).

\bibitem{miranskybook}
V.~A.~Miransky, {\it Dynamical symmetry breaking in quantum field theories}, 
(Singapore, World Scientific, 1993).

\bibitem{wsumrule}
S.~Weinberg,
Phys.\ Rev.\ Lett.\  {\bf 18} (1967), 507.

\bibitem{sumrule2} 
C. Bernard, A. Duncan, J. LoSecco, and S. Weinberg,
Phys. Rev. D {\bf 12}, 792 (1975); (1976); E. Floratos, S. Narison, and E. de
Rafael, Nucl. Phys. B {\bf 155}, 115 (1979); 
S.~Weinberg, {\it The Quantum Theory of Fields} (Cambridge Univ. Press,
Cambridge, 1996), v. 2, p. 266. 

\bibitem{svz}
M. Shifman, A. Vainshtein, and V. Zakharov, Nucl. Phys. B {\bf 147}, 385
(1979). 

\bibitem{dg}
A numerical analysis of these sum rules in QCD in terms of data on the spectral
functions for is given in J. Donoghue and E. Golowich, Phys. Rev. D {\bf 49}, 
1513 (1994). 

\bibitem{bs}
E. Salpeter and H. Bethe, Phys. Rev. {\bf 84}, 1232 (1951). 

\bibitem{nakanishi}
For an early review, see N. Nakanishi, Prog. Theor. Phys. Suppl.  {\bf 43}, 1
(1969).

\bibitem{lane74}
K. Lane, Phys. Rev. D {\bf 10}, 2605 (1974).

\bibitem{mn74}
T. Maskawa and H. Nakajima, Prog. Theor. Phys. {\bf 52}, 1326 (1974).

\bibitem{kugo}
R. Fukuda and T. Kugo, Nucl. Phys. B {\bf 117}, 250 (1974);
T. Kugo, Phys. Lett. B {\bf 76}, 625 (1978).

\bibitem{higashijima}
K.~Higashijima,
Phys.\ Rev.\ D {\bf 29}, 1228 (1984).

\bibitem{Aoki:1990aq}
K.~I.~Aoki, M.~Bando, T.~Kugo and M.~G.~Mitchard,
Prog.\ Theor.\ Phys.\  {\bf 85}, 355 (1991).

\bibitem{Kugo:1992pr}
T.~Kugo and M.~G.~Mitchard,
Phys.\ Lett.\ B {\bf 282}, 162 (1992). 

\bibitem{Bando-Harada-Kugo}
M.~Bando, M.~Harada and T.~Kugo,
Prog.\ Theor.\ Phys.\  {\bf 91}, 927 (1994). 

\bibitem{avsrev}
R. Alkofer and L. von Smekal, Phys. Rept. {\bf 353}, 281 (2001).

\bibitem{marisroberts03}
P. Maris and C. D. Roberts, Int. J. Mod. Phys. E {\bf 12}, 297 (2003).

\bibitem{gamcon}
Our gamma matrix and metric conventions are those of Bjorken and Drell. 

\end{thebibliography}
\end{document}